# A Therapeutic Stress Ball
# to Monitor Hand Dexterity and Electrodermal Activity

Authors: Fereshteh Shahmiri, Steven Schwartz, Can Usanmaz, Alexander Fache, Alec Mills.

**ECE 6781– Biomedical Sensing Systems** (Final Project) - Instructor: Dr. Omer Inan

Date: December 3rd, 2020

Executive Summary

We designed, built, and tested a triboelectric nanogenerator-based (TENG) therapeutic stress ball to provide gesture monitoring and physiological data on patients requiring physical therapy of various degrees. The device utilizes a 5-layer stack of silicone and braided silver-coated nylon rope electrodes to create a sensor network that monitors 40-points across the surface of a semi-spherical prototype. A modified version of a standard ECG circuit was utilized to provide proper loading, noise rejection, filtering, and phase of the TENG signals along with multiplexing of the many electrodes. All system components were selected with a final embedded system in mind. Testing of the device was conducted utilizing an Arduino Uno and an EVAL-AD5940BIOZ evaluation board for electrodermal activity for stress and/or pain after exercise. An accelerometer was included for device activation and hand tremor detection. Upon testing, the self-powered TENG sensors produce positive impulses upon contact and negative impulses upon release of contact from the surface of the ball. Furthermore, finger removal detection was demonstrated by capturing the associated negative impulse by maintaining the bipolar signal in our conditioning circuit. EDA results indicate silver-coated nylon as a potentially good dry-electrode which can be used with even more electrodes for bio-impedance or ECG capture to further expand the device functionality. A MATLAB-based GUI was designed to provide the user with data tracking and visual monitoring of the data via serial communication from the microcontrollers. Finally, it should be noted that this provides a means for low-cost low-power gesture tracking without the use of flexible capacitive grid arrays and provides the user with a pleasant tactile experience that one expects form a stress ball due to its unique material design.

Introduction

Hand dexterity, grip strength, and fine motor control are important to our daily routines but can be severely impacted by the development of prognoses such as Parkinson's disease, Arthritis, Carpal Tunnel Syndrome, or recovery after a stroke, surgery, or coma. Simple yet effective squeeze ball exercises, as shown in figure 1, have been shown to accelerate recovery, restore mobility, and reduce pain associated with the conditions [2, 4, 8, 14, 20, 21]. However, in many of these cases patients experience some level of pain and stress associated with the necessary therapeutic exercises required to recover lost motor control and hand mobility. Hence, there is a need to monitor both patient performance and compliance with respect to these exercises along with additional assessments of pain and stress levels.

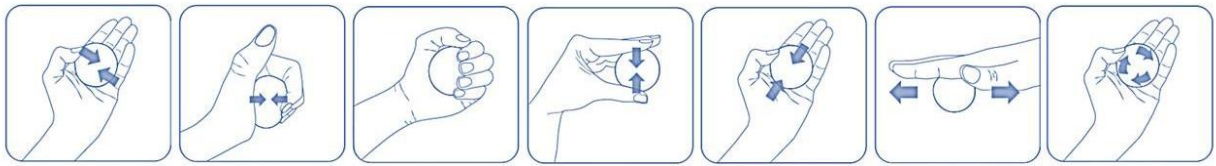

*Figure 1: Common restorative hand therapy exercises that promote strength and dexterity*

Most existing commercially available hand exercising balls do not provide quantitative data, which is crucial for assessing patient performance and pain level as well as informing clinicians for proper treatment. There are very few commercial products like [22, 23] or some research efforts like [1, 9, 15] that have incorporated pressure sensors into the device to provide grip strength data to patients. Unfortunately, these few examples, except [23], lack an adequately flexible form factor for many of the therapeutic exercises previously discussed. It must be highlighted that for post-stroke patients or those with hand arthritis and CTS, the aim is not always on grip strengthening, but on improving the mobility of fingers and different forms of grasp, while alleviating the associated pain [2, 4, 8, 17]. To the best of our knowledge, there is no single device in the market or existing research domains, with a flexible form factor that can identify the gripping patterns and correctness of the performed therapeutic exercises as well as assess the pain level that patient is experiencing before and after those exercises. Hence, we designed the first triboelectric nanogenerator-based squeezable electronic ball, known otherwise as the TENSE ball. This therapeutic device is computationally capable of addressing the discussed problems, while maintaining the innate features as its non-computational counterparts. The design of the TENSE ball involves a unique 3D-printed hollow core for the embedding of the computational hardware, material design for the creation of the triboelectric nanogenerator skin which will serve as a sensor network, and the appropriate circuitry to utilize this sensor network and integrate it with additional sensors such as an accelerometer and an electrodermal activity (EDA) sensor.

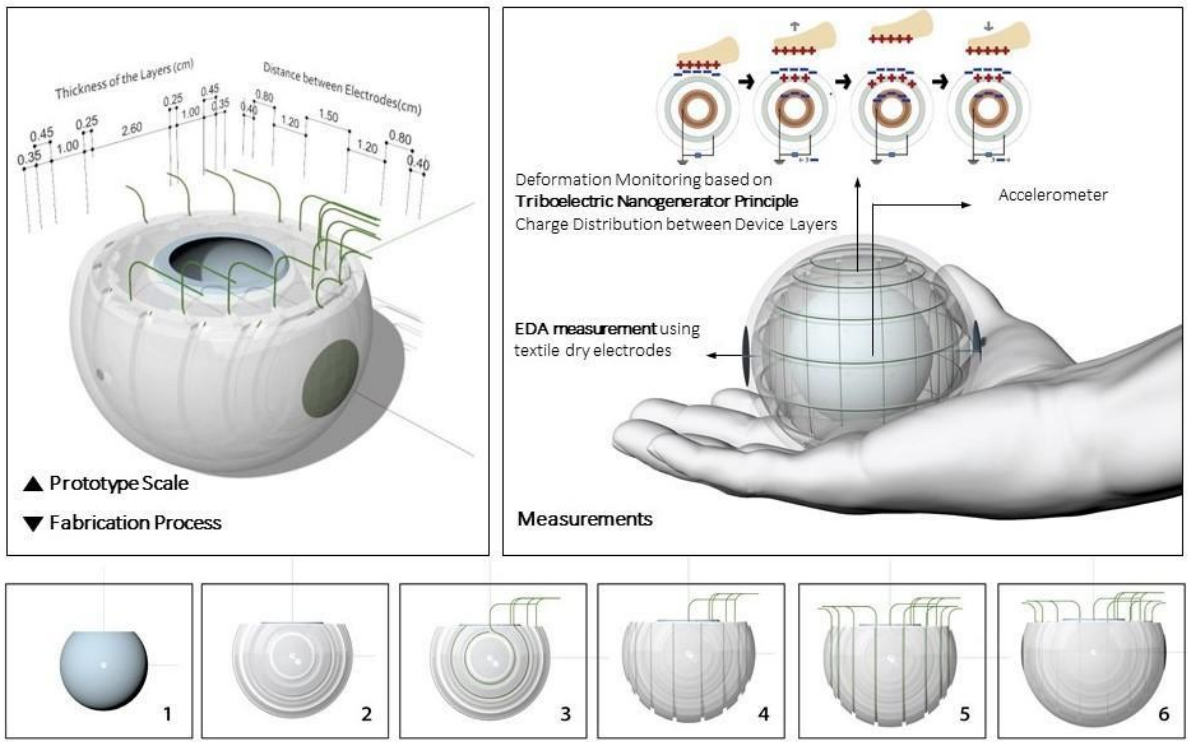

The major contributions of this design include the following. First, the capturing of psycho-physiological data utilizing an EDA sensor. Although, by itself this is not a perfect indicator of pain and stress level, it could be a valuable biomarker when used alongside the patient's subjective pain assessment as is currently used in [5, 6, 7, 12]. In the future, additional biomarkers could be captured through further integration of sensors such as ECG [11] and bioimpedance [13]. Second, it monitors the biomechanical status of hand motions and dexterity to assess the correctness of the patient's performed therapeutic exercises and can track compliance and improvement with time.

Third, it can detect hand tremors and other involuntary motion artifacts which allows for the capturing of valuable patient prognostic information while enabling error reduction and noise correction from motion artifacts which would impact accuracy of the other sensed data.

To provide the best understanding of our current system, we will integrate both the subsystem design considerations along with device construction methods and prototype testing. The TENG sensor network for gesture monitoring will first be explained as it drove the majority of the design, then the circuit considerations, the rest of the system design, the additional sensors, and finally the testing of the device and its integration with a MATLAB-based GUI.

<u>TENG Sensor Network for Gesture Monitoring</u>

To monitor the biomechanical status of hand motions, a triboelectric nanogenerator (TENG) sensor network was built via the creation of a 5-layer stack of silicone and conductive silver-coated nylon thread as seen in figure 2. Triboelectrification occurs because of charge transfer between two materials of different electronegativity [16, 18, 19]. We designed a network of electrodes in the soft body of the ball for the required charge transfer. The buildup of charges from triboelectrification results in electrostatic induction, or a charge flow observed as a current. Thus, every specific mechanical deformation of, or change in contact with, the ball generates signals and electrical power [16] which is detected as a voltage across a load resistor placed between the electrodes and ground. It should be noted that based on our tests, our system mostly responds to contact changes and varies only slightly with mechanical deformation which is advantageous for gesture recognition by minimizing noise from squeezing.

Every intersection of the top and bottom electrodes can be used as a point in the sensor network. As shown in figure 2, to encompass the whole prototype a 40-point grid will be constructed with 8 parallel geodesics of electrodes on one layer and 8 parallel geodesics on another layer perpendicular to the first. Using a dual multiplexing system each line is readout as a differential voltage with respect to ground and each point is referenced as the multiplication of the two voltages on the lines that create the intersection. A heatmap is then created for every point demonstrating where contact/motion is occurring.

The ball was constructed by first 3D printing a series of four sets of specifically designed semi-spheres as seen in Figure 2. The second and third mold sets had parallel geodesic protrusions to create trenches in the silicon for depositing of the flexible electrodes during subsequent casting. A semi-sphere was used for the prototype to enable access to the interior core for threading the EDA electrodes, and enabling easy configuration of the ends of the 16 flexible electrodes for the TENG network. The mold sets were built in sections so that they could easily be removed post-casting. The silicone mixture was comprised of a 1:1 ratio of Ecoflex 00-50 and Shore 00-50 which was mixed for 3 minutes and then injected into the molds using a large volume syringe. The electrodes were created by braiding 10 strands of Ag-coated nylon into thicker and more wieldy conductive ropes which were deposited in the trenches and secured via superglue placed only at the bottom of the semi-spheres. Stabilization of the previously assembled structure in the subsequent mold was performed using toothpicks placed through holes left in the molds to ensure that new layers are deposited equally around the exterior of the previous layer. Additionally, the first layer of silicone was chosen to be thicker than the rest to provide the user with improved tactile feel and squeezability need to compensate for the firm 3D printed core. The group members all found the resulting stress ball to be extremely enjoyable to squeeze, highlighting the results of our iterative prototyping process and material selection.

System & Circuit Design

With the final intent of the system to be embedded in the ball we chose to design our circuit using only a single supply. Furthermore, the voltage for most of the system was selected to be 3.3V based on the 3.3V voltage regulator on the microcontroller breakout board. To handle the bipolar signals produced by the TENG sensor network, we created a half-supply reference using a buffered voltage divider and used this as a rail in our breadboard along with 0V ground and 3.3V positive supply. Normally, we would have used an adder to place the bipolar TENG signal on top of the half-supply reference and then low pass filtered the DC bias out after the multiplexer but before the instrumentation amplifier, but when testing the circuit we discovered the single supply multiplexer could handle up to -0.3V signals which was enough leeway for our pre-amplified signals. So, we discarded the adder and instead the TENG network utilized two single supply, serial-controlled, analog output multiplexers to loop through the 16 flexible electrodes. A high impedance voltage divider was placed after the muxes to provide empirically verified proper loading of the TENG electrodes and a necessary path to ground for TENG-produced charge to prevent continuous charge buildup. The signal then went through RFI-protection, ESD-protection, and a single supply level-translating instrumentation amplifier with the 1.65V reference. The signal was then passively high pass filtered to remove DC offsets and baseline wander after which it was increased with a non-inverting amplifier, with a gain set to 4. The signal was then filtered through a 5$^{th}$ order lowpass filter upon which it was inverted due to the active inverting low-pass filter. To correct for this, we added a second inverting amplifier set to unity gain. All stages after the instrumentation amplifier were referenced to the 1.65V buffered half-supply reference to maintain the integrity of the bipolar signal.

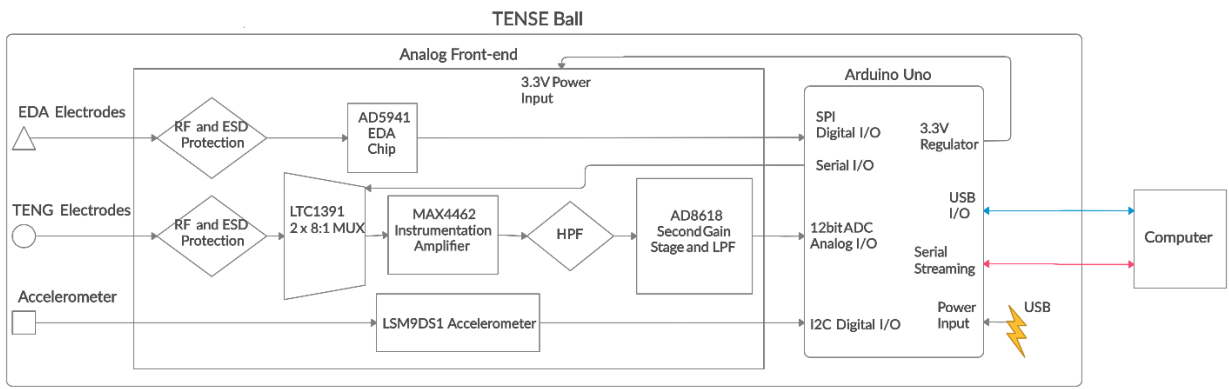

In the final system, the gain was adjusted mainly at the non-inverting amplifier to improve noise and reduce the threat of baseline wander causing signal clipping if the instrumentation amplifier's gain were too large. We settled on a gain of 4 after empirical experimentation with the circuit. For all op-amps excepting the instrumentation amplifier, the AD8618 was used due to its good single-supply operation, unity gain stability reducing the need for additional passives, and the 4 op-amps configuration in a small 14-pin TSSOP package. The instrumentation amplifier was used here to provide common mode rejection for the electrodes with respect to ground. In the prototype all exposed flexible electrodes where shrink-wrapped and taped to prevent shorting and reduce interference. The 5$^{th}$ order low pass filter's bandwidth was kept at 157Hz from the AD8220 based ECG circuit from Analog Devices, as it was a perfect bandwidth for our device which could not exceed 200Hz due to our microcontroller/codes sample rate and had frequency components varying from sub-hertz to the few hundreds of hertz.

The EDA sensor utilizes two dry electrodes constructed by the placement of a third electrode layer, made up of the similar silver-coated nylon fabric, over the top of two of the 40-points in the sensor network at opposite poles on the ball as shown in figure 2. The validity of this dry electrode for wearable-based EDA has been shown in [3, 10] and proven empirically. An AD5940 was used as the analog front end and ADC and was digitally interfaced to the MCU. The EVAL-AD5940BIOZ evaluation kit was used to interface with the EDA chip over serial connection given the difficulties of the library discrepancies between the MCU on the evaluation board and the Arduino Uno's MCU. Electrodes from the evaluation board were connected via microUSB using UART to a breadboard circuit utilizing the proper 1kOhm current limiting resistor and 15nF and & 470nF international standard isolation capacitors. An LSM9DS1 accelerometer with an integrated front end and ADC was also digitally interfaced to the MCU and the X, Y, Z acceleration values were read out after each TENG cycle.

For the embedded system an Espruino Wi-Fi will be used as the main microcontroller due to its size, affordability, Wi-Fi compatibility (instead of BLE due to data streaming capability), high analog and digital I/O count, and on-board 3.3V voltage regulator. However, for the breadboard an Arduino Uno was used due to our familiarity with the language and bugs inside the Espruino's Github provided code. To power the system, we would ideally use a small 6V lithium ion rechargeable battery, with a port on the hollow cores shell accessible via a silicone flap. However, for our testing we used laptop ground and power via microUSB. It is important to note, that the TENG signals are self-powered which helps reduce power consumption in the final device.

Testing & Data Interfacing

The envisioned final system would have an operating cycle in which the ball registers an accelerometer magnitude larger than a set threshold activating the EDA cycle. A three-minute EDA baseline would be taken by placing two fingers on the corresponding conductive fabric electrodes. Afterwards, the device would go into TENG mode in which it monitors gesture control. At the conclusion of the therapeutic exercise the user would shake the ball once again above threshold to initiate a final three-minute EDA cycle to obtain the post-exercise physiological data relating to stress/pain. The LSMD9S1 was selected due to its power efficient "always-on" power saving mode for application such as this; although, this was not yet implemented in the actual prototype.

Testing during prototype was multistaged with the first stage based on testing a flat 2D 4x4 TENG sensor network to acquire signals used in the LTSpice simulation show in figure 3. Following this sensor validation and subsequent construction of the TENSE ball semi-sphere prototype, we first tested individual electrode readouts to ensure the signal amplitude fell off with movement of the finger away from the electrode as shown in figure 4. The signal here is interesting as it is a positive impulse during skin contact and negative impulse only after the release of skin contact. This allowed us to create a bipolar colormap program for GUI purposes for showing when contact is made (due to positive signals) and disconnected (due to negative signals). To demonstrate gesture capture we made a GUI that shows the cross-points of the TENG array when multiple fingers are contacting the ball simultaneously as seen in figure 4. It should be noted that we were unable to demonstrate gesture capture and correct finger positioning, but we did get asymmetric heatmapping upon contact with the ball. When signals from all 16 electrodes were captured and analyzed using MATLAB, slight variations in height were noticeable with the distance from the ball, but mapping in both X and Y direction was not possible. To do this a pixel-based grid readout system or alternative latching-based readout method might be required.

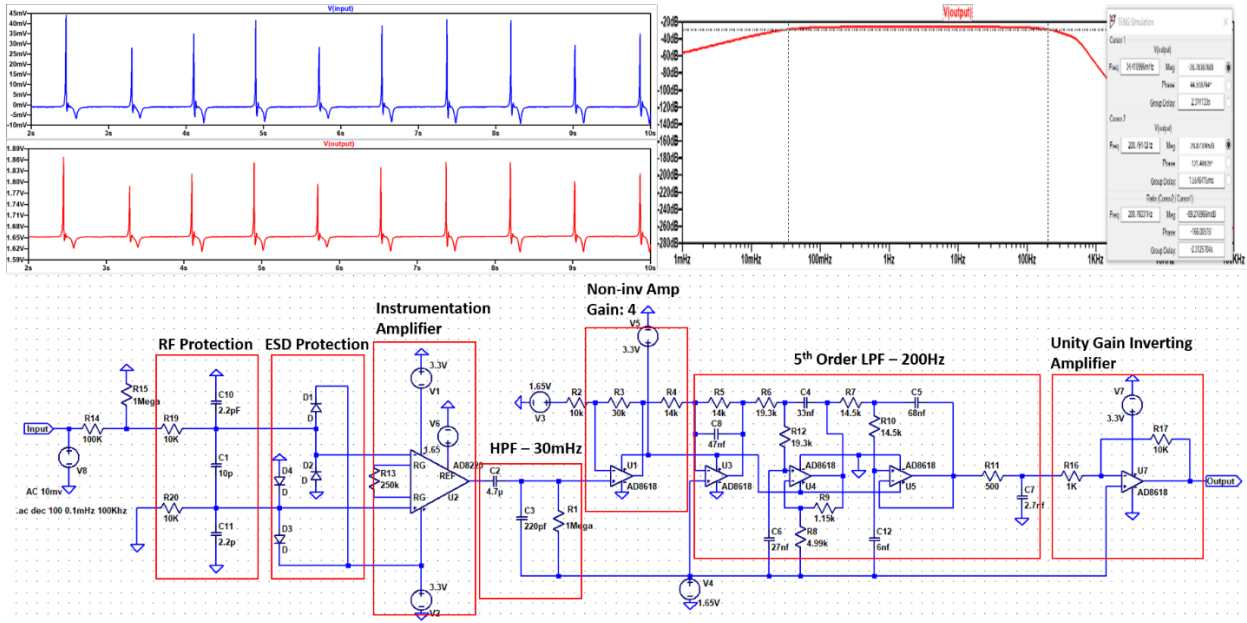

*Figure 4 – TENG Condition Circuit Schematic and LTSpice simulation. The bottom image shows a modified version of the final circuit schematic as a dual supply instrumentation amplifier was used in place of the single-supply MAX4622 in the breadboard. The top left image shows the original TENG signal of the 4x4 2D test array data. The middle left image shows the resulting signal at the output of the modeled conditioning circuit. The top right image is of the modeled frequency response of the TENG circuit.*

Prior to circuit development, an Analog Discovery 2 with 1MOhm input impedance was used to detect signal from the TENG network. It is here we realized the importance of impedance in properly loading the TENG signals and the requirement for the high impedance voltage divider (which does not actually do much division) to allow a high impedance source to ground and properly load the circuit. All data was originally captured using serial port read on Arduino as shown in figure 5 (left), but then switched to serial communication directly to the MATLAB GUI program in figure 5 (right).

The EDA circuit was tested using the evaluation kit. Two ECG electrodes were attached via microUSB utilizing UART protocol to the AD5940BIOZ board. These electrodes were connected to the breadboard to provide current limiting and isolation and the silver-coated nylon fabric was connected to the breadboard via braided conductive nylon rope stitched into the edge of the EDA electrode pads. The TIA resistor was set to automatic scaling using the software to properly size the signal for the onboard ADC, the DFT was set to 16 with a Hanning Window, and total magnitude of the impedance was readout as a timeseries via a serial connection and inverted to get the conductance. The sample rate was 4 fps and the frequency was 100Hz. The circuit was compared to its shorted version to ensure we did not have too much contact resistance from the breadboard. A sample timeseries is shown in the GUI image in the figure 5 with a variation from the shorted signal of approximately 70kOhm impedance or a 14.3 µS conductance.

Conclusion

In the end we are happy to have been able to deliver a fully assembled prototype with working circuits, peripherals, and data communication pathways. We were able to deliver well-conditioned TENG signals and get our MCU loop with the peripherals to interact with our MATLAB GUI in real-time with minimal delay. It was not until after everything was assembled and connected to the MATLAB GUI for processing that we learned of the potential flaws of our chosen electrode readout system, and of the difficulties of placing continuous contact force across a spherical surface. In the future, we believe all components could be embedded onto a shield for an Espruino module and inserted into the device to deliver wireless data transfer to create a very real and usable system. To do this we would first need to fix the readout mechanisms by either some type of calibration, improved signal processing, or TENG pixelization utilizing transistor networks, flexible readout lines, and tiny swatches of electrodes for improved surface area coverage yielding large signals for reduce power consumption. The self-powered nature of these signals provides the potential for a very real low-power embedded system that could provide improved data collection which is driving a transformation in the medical space for how patients interact and are involved in their own treatment.

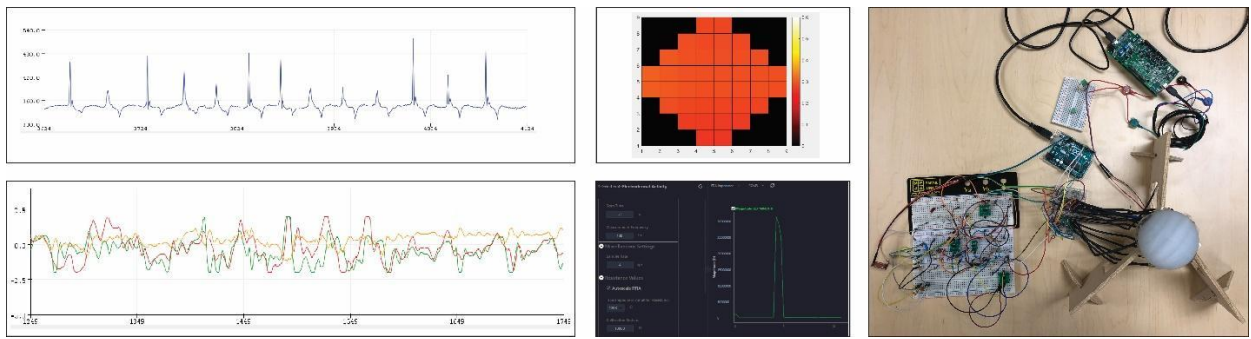

**References**

[1] Adams, Alexander T., et al. "Keppi: A tangible user interface for self-reporting pain." Proceedings of the 2018 CHI Conference on Human Factors in Computing Systems. 2018.

[2] Almel, Waseemakram. "Effectiveness of squeezing ball on reduction of pain during insertion of intravenous cannula among hospitalized preschool children in selected hospitals at Vijayapur" (2018).

[3] An, Xiang, Orathai Tangsirinaruenart, and George K. Stylios. "Investigating the performance of dry textile electrodes for wearable end-uses." *The Journal of The Textile Institute* 110.1 (2019): 151-158.

[4] Arya, Kamal Narayan, Shanta Pandian, and Dharmendra Kumar. "Task-based mirror therapy enhances ipsilesional motor functions in stroke: a pilot study." Journal of bodywork and movement therapies 21.2 (2017): 334-341.

[5] Bari, Dindar S., et al. "Electrodermal activity responses for quantitative assessment of felt pain." *Journal of Electrical Bioimpedance* 9.1 (2018): 52-58.

[6] Braithwaite, Jason J., et al. "A guide for analyzing electrodermal activity (EDA) & skin conductance responses (SCRs) for psychological experiments." *Psychophysiology* 49.1 (2013): 1017-1034.

[7] Boucsein, Wolfram. Electrodermal activity. Springer Science & Business Media, 2012.

[8] Deveza, Leticia A., et al. "Efficacy of combined conservative therapies on clinical outcomes in patients with thumb base osteoarthritis: protocol for a randomized, controlled trial (COMBO)." BMJ open 7.1 (2017).

[9] Ganeson, Suhassni, Radzi Ambar, and Muhammad Mahadi Abdul Jamil. "Design of a low-cost instrumented glove for hand rehabilitation monitoring system." 2016 6th IEEE International Conference on Control System, Computing and Engineering (ICCSCE). IEEE, 2016.

[10] Haddad, Peter A., et al. "Breathable dry silver/silver chloride electronic textile electrodes for electrodermal activity monitoring." *Biosensors* 8.3 (2018): 79.

[11] Lourenço, André, Hugo Silva, and Ana Fred. "Unveiling the biometric potential of finger-based ECG signals." *Computational intelligence and neuroscience* 2011 (2011).

[12] Naranjo-Hernández, David, Javier Reina-Tosina, and Laura M. Roa. "Sensor Technologies to Manage the Physiological Traits of Chronic Pain: A Review." *Sensors* 20.2 (2020): 365.

[13] Noh, Hyung Wook, et al. "Ratiometric impedance Sensing of fingers for Robust identity Authentication." *Scientific reports* 9.1 (2019): 1-12.

[14] Pongantung, Henny, et al. "the effect of the ball grasping therapy on the strength of upper limb muscles in post-stroke patients from Stella Maris hospital in Makassar". Urban Health 2.1 (2019).

[15] Schoenhals, William A. The Research and Development of a Wireless Hand Dynamometer as Part of a Physiological Health Monitoring System. The University of Tulsa, 2017.

[16] Shahmiri, Fereshteh, et al. "Serpentine: A self-powered reversibly deformable cord sensor for human input." Proceedings of the 2019 CHI Conference on Human Factors in Computing Systems. 2019.

[17] Ünver, Seher, and Neriman Akyolcu. "The effect of hand exercise on reducing the symptoms in hemodialysis patients with carpal tunnel syndrome." Asian journal of neurosurgery 13.1 (2018): 31.

[18] Wang, Yang, Ya Yang, and Zhong Lin Wang. "Triboelectric nanogenerators as flexible power sources." npj Flexible Electronics 1.1 (2017): 1-10.

[19] Wang, Zhong Lin. "Triboelectric nanogenerators as new energy technology and self-powered sensors–Principles, problems and perspectives." Faraday discussions 176 (2015): 447-458.

[20] Wietlisbach, Christine M. Cooper's Fundamentals of Hand Therapy E-Book: Clinical Reasoning and Treatment Guidelines for Common Diagnoses of the Upper Extremity. Elsevier Health Sciences, 2019.

[21] Wu, John Z., et al. "An evaluation of the contact forces on the fingers when squeezing a spherical rehabilitation ball." Bio-medical materials and engineering 29.5 (2018): 629-639.

[22] https://www.dynatomyproducts.com/products/varigrip-sport/variable-resistance/

[23] https://mysquegg.com/